\documentclass[showpacs,preprintnumbers,onecolumn]{revtex4}%
\usepackage{amsfonts}
\usepackage{amsmath}
\usepackage[dvips]{graphicx}
\usepackage{dcolumn}
\usepackage{bm}
\usepackage{amssymb}%
\providecommand{\U}[1]{\protect\rule{.1in}{.1in}}
\ifx\pdfoutput\relax\let\pdfoutput=\undefined\fi
\newcount\msipdfoutput
\ifx\pdfoutput\undefined\else
\ifcase\pdfoutput\else
\ifx\paperwidth\undefined\else
\ifdim\paperheight=0pt\relax\else\pdfpageheight\paperheight\fi
\ifdim\paperwidth=0pt\relax\else\pdfpagewidth\paperwidth\fi
\fi\fi\fi
\begin{document}
\preprint{ }
\title{Ankle phenomenon in the cosmic ray energy spectrum}
\author{Yukio Tomozawa}
\affiliation{Michigan Center for Theoretical Physics, Randall Laboratory of Physics,
University of Michigan, Ann Arbor, MI. 48109-1040}
\date{\today }

\begin{abstract}
The author has suggested that the knee phenomenon in the cosmic ray energy
spectrum at 3 PeV can be explained as a split between a radiation-dominated
expansion and a matter-dominated expansion of an expanding heat bath. The
model proposed in 1985, in fact, predicted that high energy cosmic rays are
emitted from AGN, massive black holes, in agreement with recent data from the
Pierre Auger Observatory. Similarly, the ankle phenomenon at 3 EeV is shown to
be explained by a split between inflational expansion and ordinary material
expansion of the expanding heat bath, not unlike that in the expansion of the
universe. All the spectral indicies in the respective regions of the energy
spectra agree with the theoretical calculation from the respective expansion rates.

\end{abstract}

\pacs{04.70.-s, 95.35.+d, 95.85.Pw, 98.54.Cm, 98.70.Sa, 98.80.-k}
\maketitle

\section{Introduction}

There has been a proposal to explain the knee energy by a series of nuclear
components of cosmic rays which respond to the galactic magnetic field, each
having an anti-Gaussian energy spectrum at different peak positions\cite{wolf}%
. This model is not only inconsistent with the observation of the nuclear
components of cosmic rays below the knee energy\cite{nuclear}, but it also
requires the remainder of the energy spectrum to have a big hole near the knee
energy and the totality of the energy spectrum to have a relatively simple
shape, with spectral indicies of 2.5 and 3.0 below and above the knee energy
of 3 PeV, respectively. It assumes too many accidental matchings of the
intensities of the different mechanisms. It is desirable to have a unified
mechanism which gives several spectral indices, so that the matching of
intensities comes about naturally. Since 1985 the author has proposed such a
model, which explains the knee energy phenomenon. In this model, the galactic
components of cosmic rays which are created by supernova explosion, in
particular, the nuclear components, are much smaller than the main component
of the cosmic ray spectrum. This assumption is consistent with observational
data for the nuclear components\cite{nuclear}. This article explains the ankle
phenomenon from an inflational expansion, as a natural extension of the 1985 model.

\section{The 1985 model and its extension}

In a series of articles\cite{cr1}-\cite{cr9} since 1985, the author has
presented a model for the emission of high energy particles from AGN. The
following is a summary of the model.

1) Quantum effects on gravity yield repulsive forces at short
distances\cite{cr1},\cite{cr3}.

2) The collapse of black holes results in explosive bounce back motion with
the emission of high energy particles.

3) Consideration of the Penrose diagram eliminates the horizon problem for
black holes\cite{cr4}. Black holes are not black any more.

4) The cause of supernova explosion after gravitational collapse of massive
stars can be reduced to these repulsive forces.

5) The knee energy for high energy cosmic rays can be understood as a split
between a radiation-dominated expansion and a matter-dominated expansion, not
unlike that in the expansion of the universe. (See page 10 of the lecture
notes\cite{cr3} and the next section.)

6) Neutrinos and gamma rays as well as cosmic rays should have the same
spectral index for each AGN, active galactic nucleus. They should show a knee
energy phenomenon, a break in the energy spectral index at 3 PeV, similar to
that for the cosmic ray energy spectrum.

It is worthwhile to mention that this model has been supported by recent data
from the Pierre Auger Observatory\cite{auger}, which has found a possible
correlation between the sources of high energy cosmic rays and AGN.

Further discussion of the knee energy in the model yields the existence of a
new mass scale in the knee energy range, in order to have the knee energy
phenomenon in the cosmic ray spectrum\cite{crnew}. The following are
additional features of the model.

7) The proposed new particle with mass in the knee energy range (at 3 PeV) may
not be stable, as in the case of the standard model. The standard model has
particles at the 100 GeV mass scale, such as W and Z bosons, which are
unstable. If it is a member of a supersymmetric multiplet and weakly
interacting with ordinary particles, the stable particle of lowest mass
becomes a candidate for a dark matter particle (DMP). The only requirement is
that such particles must be present in AGN or black holes so that the knee
energy phenomenon is observed when cosmic rays are emitted from AGN.

8) Using the supersymmetric theory of GLMR-RS (Giudice-Luty-Murayama-Rattazzi;
Randall-Sundrum)\cite{glmr}, \cite{rs}, the lowest mass corresponding to a
knee energy mass of 3 PeV is 8.1 TeV. It is shown that the sum of 8 gamma ray
observations from unknown sources in a HESS data\cite{hess2} has a definite
peak at 7.6 $\pm$ 0.1 TeV\cite{hessanal1}, \cite{hessanal2},\cite{hessanal3}.

9) There are several other particles with mass between 8 TeV \ and 3 PeV in
the GLMR-RS theory.

We assume that the target mass range for the search is from 8 TeV to 3 Pev. In
particular, the 3 PeV target is of prime importance, since it is a starting
point for the discovery of a new mass scale. Moreover, it provides bases for
matter-dominated expansion of black holes and possibly of the universe. It is
expected that such particles are produced abundantly in the process of cosmic
ray production from AGN as well as in the process of universe expansion. But
most particles produced in AGN decay. They are produced in pairs in high
energy cosmic ray showers and subsequently decay.

10) The production of a 3 PeV particle in the atmosphere requires the incident
energy of cosmic rays or DMP at the GZK cutoff range\cite{gzk}. The intensity
is so small that one does not expect production in the atmosphere. The only
possibility for such a particle is by a DMP target. With a heavy mass of 8 TeV
for DMP, one can reduce the incident energy to 5.6*10$^{17}$ eV, so that it
has a higher intensity.

11) The measurement of a 3 PeV bump amounts to the measurement of the DMP
distribution. From the muon energy spectrum one can get the DMP distribution
in the 10$^{12}$ cm range and from the neutrino energy spectrum one can get
the DMP distribution for the whole galaxy. This provides an important tool for
the study of dark matter physics\cite{hessanal3}.

One may call the proposed new particle at 3 PeV the Cion. This is an acronym
for Cosmic Interface Particle. It is also taken from the Chinese word for
knee, Xi \ (pronounced as shi).

\section{Spectral index and expansion rate}

As described in the previous section, the model started from the realization
that quantum effects on gravity yield a repulsive force at short distances. As
a result, the collapse of a black hole proceeds to an explosion and an
expanding heat bath emits various kinds of particles. This is the reason for
the emission of high energy cosmic rays, gamma rays, neutrinos and DMP from
black holes. The number of an emitted particle $X$ \ with spin $s$ is
calculated by\cite{cr1},\cite{cr2},\cite{cr3}%
\begin{equation}
N_{X}(E)=\frac{2s+1}{2\pi^{2}}E^{2}\int\frac{4\pi R^{2}dt}{e^{E/kT-\mu/kT}%
\pm1},
\end{equation}
where $R$ is the Radius of the heat bath with temperature $T$\ \ that emits
particles and $E^{2}$ is a phase space factor. The $\pm$ sign in the
denominator is for fermions/bosons. With the assumption of the expansion rate%
\begin{equation}
t=bR^{\alpha}%
\end{equation}
and%
\begin{equation}
R=\frac{l}{kT},
\end{equation}
where b, l and $\alpha$ are constants, one can compute the number of particles%
\begin{equation}
N_{X}(E)=\frac{A_{X,\alpha}}{E^{\alpha}},
\end{equation}
where%
\begin{equation}
A_{X,\alpha}=\frac{2(2s+1)b\alpha l^{\alpha+2}}{\pi}\int_{0}^{\infty}%
\frac{x^{\alpha+1}dx}{e^{x-\mu_{0}x}\pm1}%
\end{equation}
and%
\begin{equation}
\mu_{0}=\mu/E,\text{ \ \ }x=E/kT\text{,}%
\end{equation}
where $\mu_{0}$ can be put to zero for high energy $E$.

Then, the differential energy flux is given by%
\begin{equation}
f_{X}(E)=\mid\frac{dN_{X}(E)}{dE}\mid=\frac{\alpha A_{X,\alpha}}{E^{\alpha+1}%
}. \label{spec}%
\end{equation}
This is the energy spectrum of cosmic rays.

\section{The knee energy and a new mass scale}

From the expansion rate in cosmology, the exponent $\alpha$ can be estimated
as%
\begin{equation}
\alpha=2\text{ \ \ \ \ \ \ \ \ radiation-dominated regime}%
\end{equation}
and%
\begin{equation}
\alpha=3/2\text{ \ \ \ \ \ \ \ \ \ matter-dominated regime.}%
\end{equation}
This gives the cosmic ray energy spectrum above the knee energy%
\begin{equation}
f_{X}(E)\approx1/E^{3}%
\end{equation}
and below the knee energy%
\begin{equation}
f_{X}(E)\approx1/E^{2.5},
\end{equation}
using Eq. (\ref{spec}).

This is exactly the observed spectrum of cosmic rays. That is the explanation
for the observed spectrum and the existence of the knee energy, proposed in my
model in 1985\cite{cr1}..\cite{cr9}. More recently, it was
realized\cite{crnew} that the model requires the existence of a mass scale at
3 PeV in order to produce the knee energy phenomenon at 3PeV, since without it
all ordinary particles behave as massless radiation at temperature 3 PeV. The
existence of a new mass scale is the starting point for the discussion of a
DMP and its family in the previous work\cite{hessanal1},\cite{hessanal2}%
,\cite{hessanal3}.

\section{The ankle phenomenon}

The basic tenet of the 1985 model is a close connection between the expansion
in cosmology and a heat bath expansion after gravitational collapse in a black
hole. Then, it is quite natural to think that the material expansions
(radiation-dominated and matter-dominated) are preceded by an inflational
expansion. For an inflational expansion%
\begin{equation}
R=Ae^{\lambda t}%
\end{equation}
or%
\begin{equation}
t=\frac{\ln(R/A)}{\lambda}%
\end{equation}
and then%
\begin{equation}
dt=\frac{dR}{\lambda R}%
\end{equation}
hence from Eq. (\ref{spec}) one gets for $\alpha$ = 0%
\begin{equation}
f_{X}(E)\approx0.
\end{equation}

However, at higher energy E, a linear expansion dominates and hence%
\begin{equation}
R=A(1+\lambda t)
\end{equation}
and for $\alpha$ = 1 in Eq. (\ref{spec}) one gets%
\begin{equation}
f_{X}(E)\approx1/E^{2}%
\end{equation}
for high energy E. This is consistent with the observed spectrum above the
ankle energy of 3 EeV.

For the expansion%
\begin{equation}
R=A\sum\frac{(\lambda t)^{n}}{n!},
\end{equation}
if%
\begin{equation}
R\approx\frac{t^{n}}{n!}%
\end{equation}
dominates, the spectrum becomes%
\begin{equation}
f_{X}(E)\approx\frac{n!^{1/n}}{n^{2}E^{1+1/n}} \label{inflation}%
\end{equation}
Since by Stirling formula%
\begin{equation}
\frac{n!^{1/n}}{n^{2}}<\frac{e^{1/n-1}n^{1/2n}}{n} \label{stirling}%
\end{equation}
the right hand side of Eq. (\ref{stirling}) vanishes for large n, Eq.
(\ref{inflation}) vanishes for large n. This explains the transition in the
energy spectrum from $1/E^{2}$ at high energy to $0$ at low energy in the
inflational expansion.

\section{The observed spectral indices near the ankle energy}

The most recent determinations of spectral indices near the ankle energy for
the cosmic ray spectrum are\cite{cr100}%
\begin{align}
&  -2.63\pm0.02\text{ \ \ \ \ \ \ \ \ \ \ }from\text{ }the\text{ }Pierre\text{
}Auger\text{ }Observatory\\
&  -2.68\pm0.04\text{ \ \ \ \ \ \ \ \ \ \ }from\text{ }the\text{ }Hires\text{
}Group
\end{align}
above the ankle energy, and%
\begin{align}
&  -3.27\pm0.01\text{ \ \ \ \ \ \ \ \ \ \ }from\text{ }the\text{ }Pierre\text{
}Auger\text{ }Observatory\\
&  -3.33\pm0.04\text{ \ \ \ \ \ \ \ \ \ \ }from\text{ }the\text{ }Hires\text{
}Group
\end{align}
below the ankle energy. These values deviate from the predicted values of -2.0
and -3.0 respectively. These deviation may be the result of interactions of
high energy cosmic rays with cmb (Cosmic Background Radiation) photons. In
fact, such interactions caused the GZK cut-off above 10$^{20}$ eV. Therefore,
one may identify the theoretical prediction with the spectral indices for the
energy spectra of neutrinos and DMP from AGN, where little interactions are
expected with the cmb photons. Observation of the former is expected from the
Ice-Cube and Anteras neutrino detectors in the near future.

An important question, then, is whether interactions between the high energy
cosmic rays and the cmb photon can create the shifts in spectral indices%
\begin{equation}
-2.0\rightarrow-2.6
\end{equation}
above the ankle energy, and%
\begin{equation}
-3.0\rightarrow-3.3
\end{equation}
below the ankle energy. From a drastic change of the spectrum above the GZK
cut-off energy, these shifts in the spectral indices seem to be conceivable,
along with relative order of the change. It is important to resolve this question.

\section{Connection between the knee and the ankle}

This is nothing but the connection between the inflational expansion and the
material expansion, which at best is yet to be clarified for the expansion of
the universe\cite{inflation}. The advantage in the cosmic ray energy spectrum
is that it is directly observable so that any proposed model of inflation can
be tested in the cosmic ray data to some extent. In conjunction with
cosmological observations, one may hope to clarify the scenarios for inflation
from the cosmic ray data, too, in the future.

The first question is what is the significance of the ankle energy of 3 EeV.
This value is a reflection of the fractional contributions of the two
expansions, inflational and material. The higher the contribution of the
material relative to that of the inflational expansion, the higher the value
of the ankle energy. From the existing cosmic ray energy data, one may discuss
the following four cases for the connection between the two expansions.

I) Regional division of the two expansions

One may assume that there is a divisional separation of the inflational
expansion and the material expansion. In other words, while the outer region
starts inflational expansion, the inner part continues material expansion.

II) Reheating after the inflational period

This is a typical scenario after inflation in cosmology\cite{inflation}. The
temperature drops from inflational expansion can be reheated to a temperature
above the knee energy (3PeV). The outcomes of models I and II are similar. The
only difference between the two is that a larger time lag between the arrival
of cosmic rays above and below the ankle energy is required for II than for I.
If the difference in arrival time can be measured statistically, then one may
be able to differentiate the two models observationally.

III) Abortion of inflation after linear expansion

Since the low energy extension of the cosmic ray energy spectrum for
inflational expansion below the ankle energy is masked by that for material
expansion, one may assume that the inflational expansion is terminated or
aborted after a linear expansion. One has to inquire whether or not this is a
possible theoretical scenario.

IV) Separation of sources for inflational and material expansion

Depending on the gravitational collapse conditions in the black hole, one may
get dominance of either inflational or material expansion. In this case, one
should observe a separation of the sources for the two types of expansion. One
needs to observe the cosmic ray spectrum for a long period to separate the sources.

By clarifying various options for cosmic ray data which are discussed in this
section, one may hope to find a link between cosmic ray physics and cosmology.
Cases I, III and IV eliminate the necessity of reheating after inflation.

\begin{acknowledgments}
\bigskip The author would like to thank David N. Williams for reading the manuscript.
\end{acknowledgments}

\end{document}